\documentclass[11pt,onecolumn,amssymb,nofootinbib]{revtex4}
\usepackage{amsmath, amsthm, amscd, amssymb}
\usepackage{bm}
\usepackage{bbm}
\usepackage{slashed}

\begin{document}

\title{\bf Analysis of an $SU(8)$  model with a spin-$\frac{1}{2}$ field directly coupled to a gauged Rarita-Schwinger spin-$\frac{3}{2}$ field}

\author{Stephen L. Adler}
\email{adler@ias.edu} \affiliation{Institute for Advanced Study,
Einstein Drive, Princeton, NJ 08540, USA.}

\begin{abstract}
 In earlier work we analyzed  an abelianized model in which a gauged Rarita-Schwinger spin-$\frac{3}{2}$ field is directly coupled to a spin-$\frac{1}{2}$ field.  Here we extend this analysis to the gauged $SU(8)$ model for which the abelianized model was a simplified substitute.  We calculate the gauge anomaly, show that anomaly cancellation requires adding an additional left chiral  representation $\overline{8}$ spin-$\frac{1}{2}$ fermion to the original fermion complement of the $SU(8)$ model, and give options for restoring boson-fermion balance.  We conclude with a summary of attractive features of the reformulated $SU(8)$ model, including a possible connection to the
 $E_8$ root lattice.

\end{abstract}

\maketitle

\section{Introduction}

In earlier papers by the author \cite{coupl},  by the author with Henneaux and Pais \cite{henn}, and with Pais \cite{pais}, we analyzed an abelianized gauged model in which a spin-$\frac{3}{2}$ field is directly coupled to a spin-$\frac{1}{2}$ field, and carried out a perturbative calculation of the gauge anomaly.   This calculation showed that
 the combined anomaly of the
coupled spin-$\frac{3}{2}$ and spin-$\frac{1}{2}$ fields is 5 times the standard spin-$\frac{1}{2}$ anomaly, whereas prior anomaly lore \cite{lore} for spin-$\frac{3}{2}$ fields would have suggested a multiplier
of 4.  The aim of this paper is to extend the earlier calculation of \cite{coupl} to the full
$SU(8)$ model from which the abelianized model was abstracted.  We again find an anomaly multiplier
of 5 for the coupled spin-$\frac{3}{2}$ -- spin-$\frac{1}{2}$ sector of the theory.  Anomaly cancellation
in the $SU(8)$ model then requires appending an additional $\overline{8}$ spin-$\frac{1}{2}$ fermion field,
with restoration of boson-fermion balance requiring adding a corresponding number of boson fields.  This gives an updated version of the $SU(8)$ model which we originally proposed in \cite{adler1}, with interesting features which we analyze.

This paper is organized as follows.  In Sec. 2 we give the Lagrangian for the coupled sector of the
$SU(8)$ model, and formulate the momentum space equation for the fermion propagator matrix.  In Sec. 3 we find
an exact solution for the propagator and discuss its properties.  In Sec. 4 we give the vertex part
Feynman rules and  the Ward identities relating the propagator and vertex parts. In Sec. 5 we use the
computation of \cite{coupl} to calculate the anomaly in the non-Abelian coupled model.  In Sec. 6 we
give the extension of the model needed to restore $SU(8)$ anomaly cancellation and boson-fermion balance, and in Sec. 7 we
discuss features of the revised model.

\section{The coupled sector Lagrangian after $SU(8)$ symmetry breaking and the fermion propagator equation}
\subsection{The Lagrangian}
For the abelianized model of \cite{coupl}, the Lagrangian density was
\begin{align}\label{covariant}
S=&S(\psi_{\mu})+S(\lambda)+S_{\rm interaction}~~~,\cr
S(\psi_{\mu})=&\int d^4x \overline{\psi}_{\mu} R^{\mu}~~~,\cr
R^{\mu}=&i\epsilon^{\mu\eta\nu\rho}\gamma_5\gamma_{\eta}D_{\nu}\psi_{\rho}
=-\gamma^{\mu\nu\rho}D_{\nu}\psi_{\rho}~~~,\cr
D_{\nu}\psi_{\rho}=&(\partial_{\nu}+gA_{\nu})\psi_{\rho}~~~,\cr
\overline{\psi}_{\mu}=&\psi^{\dagger}_{\mu} i \gamma^0~~~\cr
S(\lambda)=&-\int d^4x \overline{\lambda} \gamma^{\nu}D_{\nu}\lambda~~~,\cr
D_{\nu}\lambda=&(\partial_{\nu}+gA_{\nu})\lambda~~~,\cr
\overline{\lambda}=&\lambda^{\dagger} i \gamma^0~~~,\cr
S_{\rm interaction}=&m\int d^4x(\overline{\lambda}\gamma^{\nu}\psi_{\nu}-
\overline{\psi}_{\nu}\gamma^{\nu}\lambda)~~~.
\end{align}
We now generalize back to the $SU(8)$ model from which Eq. \eqref{covariant} was abstracted,
by inserting $SU(8)$ group indices, and by replacing the mass $m$ in the abelianized coupling term
by the classical $SU(8)$ symmetry-breaking minimum $\overline{\phi}^{[\alpha \beta \gamma]}$ of the 56 representation scalar field
$\phi^{[\alpha \beta \gamma]}$ (with the interaction coupling absorbed into $\overline{\phi}$).  So our starting point for the analysis of this paper is
\begin{align}\label{covariant1}
S=&S(\psi_{\mu}^\alpha)+S(\lambda_{[\alpha\beta]})+S_{\rm interaction}~~~,\cr
S(\psi_{\mu}^\alpha)=&\int d^4x \overline{\psi}_{\mu\,\alpha} R^{\mu\,\alpha}~~~,\cr
R^{\mu\,\alpha}=&i\epsilon^{\mu\eta\nu\rho}\gamma_5\gamma_{\eta}D_{\nu}\psi_{\rho}^\alpha
=-\gamma^{\mu\nu\rho}D_{\nu}\psi_{\rho}^\alpha~~~,\cr
D_{\nu}\psi_{\rho}^\alpha=&\partial_{\nu}\psi_{\rho}^\alpha+gA_{\nu}^A t_{A\,\delta}^\alpha \psi_{\rho}^\delta~~~,\cr
\overline{\psi}_{\mu\,\alpha}=&(\psi_{\mu}^\alpha)^{\dagger} i \gamma^0~~~\cr
S(\lambda_{[\alpha\beta]})=&-\int d^4x \overline{\lambda}^{[\alpha\beta]} \gamma^{\nu}D_{\nu}\lambda_{[\alpha\beta]}~~~,\cr
D_{\nu}\lambda_{[\alpha\beta]}=&\partial_{\nu}\lambda_{[\alpha\beta]}+gA_{\nu}^AX_{A[\alpha\beta]}^{[\gamma\delta]}
\lambda_{[\gamma\delta]}~~~,\cr
\overline{\lambda}^{[\alpha\beta]}=&(\lambda_{[\alpha\beta]})^{\dagger} i \gamma^0~~~,\cr
X_{A[\alpha\beta]}^{[\gamma\delta]}=&\frac{1}{2}(t_{A\,\alpha}^\gamma \delta^\delta_\beta
-t_{A\,\beta}^\gamma \delta^\delta_\alpha-t_{A\,\alpha}^\delta \delta^\gamma_\beta
+t_{A\,\beta}^\delta \delta^\gamma_\alpha    )~~~\cr
S_{\rm interaction}=&\int d^4x\big(\overline{\lambda}^{[\alpha\beta]} \gamma^{\nu} \psi_{\nu}^{\gamma} \overline{\phi}^{\,*}_{[\alpha \beta \gamma]}
-\overline{\psi}_{\nu\gamma}\gamma^{\nu}\lambda_{[\alpha\beta]}\overline{\phi}^{[\alpha\beta\gamma]}\big)~~~,
\end{align}
where $t_{A\,\delta}^\alpha $ is the $SU(8)$ generator for the fundamental 8 representation.

Introducing Fourier transforms
\begin{align}\label{fourier1}
\psi_{\mu}^\alpha(x)=&\frac{1}{(2\pi)^4} \int d^4k e^{ik \cdot x} \psi_{\mu}^\alpha[k]~~~,\cr
\lambda_{[\alpha\beta]}(x)=&\frac{1}{(2\pi)^4} \int d^4k e^{ik \cdot x} \lambda_{[\alpha\beta]}[k]~~~,\cr
\end{align}
the action of Eq. \eqref{covariant} takes the form (with $\slashed{k}=\gamma^{\nu}k_{\nu}$, and relabeling dummy indices in the interaction term)
\begin{align}\label{momaction}
S=& \frac{1}{(2\pi)^4} \int d^4k  S[k]~~~,\cr
S[k]=& -i \overline{\psi}_{\mu\,\gamma}[k]\gamma^{\mu\nu\rho} k_{\nu}\psi_{\rho}^\gamma[k]-i\overline{\lambda}^{[\alpha\beta]}[k]\slashed{k}\lambda_{[\alpha\beta]}[k]
+\big(\overline{\lambda}^{[\alpha\beta]}[k]\gamma^{\rho}\psi_{\rho}^\gamma[k]\overline{\phi}^*_{[\alpha\beta\gamma]}-  \overline{\psi}_{\mu\,\gamma}[k]\gamma^{\mu}\lambda_{[\alpha\beta]}[k]\overline{\phi}^{[\alpha\beta\gamma]}\big)~~~\cr
=&\big( \overline{\psi}_{\mu\,\gamma}[k]~~\overline{\lambda}^{[\alpha\beta]}[k]\big) {\cal M}\left[\begin{array}{c}
 ~\gamma~~~\,\epsilon\kappa\\ \alpha\beta ~~\delta\\\end{array} \right] \left( \begin{array} {c}
 \psi_{\rho}^\delta[k]  \\
 \lambda_{[\epsilon\kappa]}[k] \\  \end{array}\right)~~~.\cr
\end{align}
Here ${\cal M}\left[\begin{array}{c}
 ~\gamma~~~\,\epsilon\kappa\\ \alpha\beta ~~\delta\\\end{array} \right] $ is the matrix
\begin{equation}\label{matrixM}
{\cal M}\left[\begin{array}{c}
 ~\gamma~~~\,\epsilon\kappa\\ \alpha\beta ~~\delta\\\end{array} \right] = \left( \begin{array} {c c }
-i  \gamma^{\mu\nu\rho} k_{\nu} \delta^\gamma_\delta  & -\gamma^{\mu} \overline{\phi}^{[\epsilon\kappa\gamma]}   \\
  \gamma^{\rho} \overline{\phi}^*_{[\alpha\beta\delta]}    &  -i\slashed{ k}\delta^{[\epsilon\kappa]}_{[\alpha\beta]} \\  \end{array}\right)~~~,
\end{equation}
where we have defined
\begin{equation}\label{doubledelta}
\delta^{[\epsilon\kappa]}_{[\alpha\beta]}=\frac{1}{2}\Big( \delta^\epsilon_\alpha \delta^\kappa_\beta
-\delta^\kappa_\alpha\delta^\epsilon_\beta\Big)~~~.
\end{equation}

The propagator for the coupled $\psi_{\mu}^\alpha$ and $\lambda_{[\alpha\beta]}$ fields is the matrix ${\cal N}\left[\begin{array}{c}
 ~\delta~~~\,\alpha^\prime\beta^\prime\\ \epsilon\kappa~~\gamma^\prime\\\end{array} \right]$ that is inverse to ${\cal M}$,
\begin{align}\label{Minverse}
{\cal M}{\cal N}=&  \left( \begin{array} {c c }
\delta^{\mu}_{\sigma} \delta^{\gamma}_{\gamma^\prime} &0    \\
 0  & \delta^{[\alpha^\prime\beta^\prime]}_{[\alpha\beta]} \\  \end{array}\right)~~~.\cr
\end{align}

\section{Explicit solution for the propagator and its properties}

Writing ${\cal N}\left[\begin{array}{c}
 ~\delta~~~\,\alpha^\prime\beta^\prime\\ \epsilon\kappa~~\gamma^\prime\\\end{array} \right]$ in the form
\begin{align}\label{Minva}
{\cal N}\left[\begin{array}{c}
 ~\delta~~~\,\alpha^\prime\beta^\prime\\ \epsilon\kappa~~\gamma^\prime\\\end{array} \right]
 =&\left( \begin{array} {c c }
N_{1\rho\sigma\,\gamma^\prime}^\delta & N_{2\rho}^{[\alpha^\prime\beta^\prime\delta]}    \\
N_{3\sigma\,[\epsilon\kappa\gamma^\prime]}    &N_{4[\epsilon\kappa]}^{[\alpha^\prime\beta^\prime]}\\  \end{array} \right) ~~~,\cr
\end{align}
Eq. \eqref{Minverse} gives a set of four equations for $N_1,..., N_4$.  In terms of the matrix
\begin{equation}\label{Mdef}
M_\delta^\gamma= \overline{\phi}^{[\alpha\beta\gamma]}\overline{\phi}^*_{[\alpha\beta \delta]}~~~,
\end{equation}
which we assume to be invertible, these equations can be solved to give the unique answer
\begin{align}\label{Nformula}
N_{1\rho\sigma\,\gamma^\prime}^\delta=&\tilde{N}_{\rho\sigma}\delta^{\delta}_{\gamma^\prime} -\frac{i}{k^2}k_\rho k_\sigma \slashed{k}(M^{-1})^{\delta}_{\gamma^\prime}~~~,\cr
\tilde{N}_{\rho\sigma}=& \frac{-i}{2 k^2}\left[\gamma_{\sigma} \slashed{k} \gamma_{\rho}-\frac{4}{k^2}k_{\rho}k_{\sigma}\slashed{k}\right]~~~,\cr
N_{2\rho}^{[\alpha^\prime\beta^\prime\delta]}=&\frac{\slashed{k}k_\rho}{k^2} (M^{-1})^{\delta}_{\gamma} \overline{\phi}^{[\alpha^\prime \beta^\prime \gamma]}~~~,\cr
N_{3\sigma\,[\epsilon\kappa\gamma^\prime]}=&-\frac{\slashed{k}k_\sigma}{k^2} (M^{-1})^{\delta}_{\gamma^\prime}
\overline{\phi}^*_{[\epsilon\kappa\delta]}~~~,\cr
N_{4[\epsilon\kappa]}^{[\alpha^\prime\beta^\prime]}=&\frac{i \slashed{k}}{k^2} P^{[\alpha^\prime \beta^\prime]}_{[\alpha\beta]}~~~,\cr
P^{[\alpha^\prime \beta^\prime]}_{[\alpha\beta]}=&\delta^{[\alpha^\prime \beta^\prime]}_{[\alpha\beta]}
-\overline{\phi}^*_{[\alpha\beta\delta]}(M^{-1})_\gamma^\delta \overline{\phi}^{[\alpha^\prime\beta^\prime\gamma]}~~~.\cr
\end{align}

We consider in this paper only the case when $M$ is invertible, which will be illustrated with a concrete example below; when $M$ is not invertible, the action of Eq. \eqref{covariant1} has a fermionic gauge invariance that requires adding gauge fixing and ghost terms to the action.  We note that when $\overline{\phi}$ is scaled in magnitude to infinity, as
is relevant for the anomaly calculation below, the off-diagonal matrix elements of ${\cal N}$ vanish, and $N_{1\rho\sigma\,\gamma^\prime}^\delta$ reduces to its first term $\tilde{N}_{\rho\sigma}\delta^{\delta}_{\gamma^\prime}$. So in the large $\overline{\phi}$ limit, we have
\begin{align}\label{Minva1}
{\cal N}\left[\begin{array}{c}
 ~\delta~~~\,\alpha^\prime\beta^\prime\\ \epsilon\kappa~~\gamma^\prime\\\end{array} \right]
 \simeq &\left( \begin{array} {c c }
\tilde{N}_{\rho\sigma}\delta^{\delta}_{\gamma^\prime} & 0    \\
0    &\frac{i \slashed{k}}{k^2} P^{[\alpha^\prime \beta^\prime]}_{[\epsilon\kappa]}\\  \end{array} \right) \cr
=&  \left( \begin{array} {c c }
\tilde{N}_{\rho\sigma} & 0    \\
0    &\frac{i \slashed{k}}{k^2}\\  \end{array} \right) \times
\left( \begin{array} {c c }
\delta^{\delta}_{\gamma^\prime} & 0    \\
0    & P^{[\alpha^\prime \beta^\prime]}_{[\epsilon\kappa]}\\  \end{array} \right)~~~, \cr
\end{align}
which is a direct product of a Dirac gamma matrix part and an internal symmetry part.

We turn next to the properties of the matrices $M$ and $P$.  Taking the complex conjugate of $M$, we have
\begin{equation}\label{conjg}
M_\delta^{*\,\gamma}= \overline{\phi}^*_{[\alpha\beta\gamma]}\overline{\phi}^{[\alpha\beta \delta]}
=M_\gamma^\delta~~~,
\end{equation}
so $M$ is self-adjoint and when diagonalized has real eigenvalues.  Contracting $P$ with itself, we find that
\begin{equation}\label{projector}
P_{[\alpha\beta]}^{[\alpha^\prime\beta^\prime]}P_{[\alpha^\prime\beta^\prime]}^{[\mu\nu]}=P_{[\alpha\beta]}^{[\mu\nu]}~~~,
\end{equation}
so $P$ is a projector.  Contracting $P_{[\alpha\beta]}^{[\alpha^\prime\beta^\prime]}$ on the right with $\overline{\phi}_{[\alpha^\prime\beta^\prime\epsilon]}$, or on the left with $\overline{\phi}^{[\alpha\beta\epsilon]}$, we get
\begin{align}\label{projonphi}
P_{[\alpha\beta]}^{[\alpha^\prime\beta^\prime]}\overline{\phi}^*_{[\alpha^\prime\beta^\prime\epsilon]}=&0~~~,\cr
\overline{\phi}^{[\alpha\beta\epsilon]}P_{[\alpha\beta]}^{[\alpha^\prime\beta^\prime]}=&0~~~;\cr
\end{align}
thus we see that $P$ is a projector on the $\overline{28}$ subspace orthogonal to the scalar field classical value
$\overline{\phi}^{[\alpha\beta\epsilon]}$.

\section{Vertex parts and Ward identities}

From the action of Eq. \eqref{covariant1}, we see that the vertex part Feynman rule for the vector gauge field $A_{\nu}^A$ (omitting factors $-ig$) is
\begin{align}\label{vectorvertex}
{\cal V}^\nu_A= &\left( \begin{array} {c c }
  \gamma^{\mu\nu\rho}   t_{A\,\delta}^\alpha & 0    \\
0    &  \gamma^{\nu}X_{A[\alpha\beta]}^{[\gamma\delta]} \\  \end{array}\right)\cr
=&\Gamma_A \left( \begin{array} {c c }
  \gamma^{\mu\nu\rho}  & 0    \\
0    &  \gamma^{\nu}\\  \end{array}\right) = \left( \begin{array} {c c }
  \gamma^{\mu\nu\rho}  & 0    \\
0    &  \gamma^{\nu}\\  \end{array}\right) \Gamma_A~~~,\cr
\end{align}
with $\Gamma_A$ the internal symmetry matrix
\begin{equation}\label{gammaAdef}
\Gamma_A=
\left( \begin{array} {c c }
t_{A\,\delta}^\alpha  & 0  \\
     0 & X_{A[\alpha\beta]}^{[\gamma\delta]}\\  \end{array}\right)~~~,
\end{equation}
and with the free indices
 understood to act on the corresponding indices in row and column vectors standing
to the left and right of ${\cal V}^{\nu}_A$  in the
same matter as these indices act in Eqs. \eqref{momaction} and \eqref{matrixM}.
The corresponding axial-vector vertex part is obtained by right multiplication by $\gamma_5$,
\begin{align}\label{axialvertex}
{\cal A^{\nu}}_A= &\left( \begin{array} {c c }
 \gamma^{\mu\nu\rho} \gamma_5  t_{A\,\delta}^\alpha   & 0    \\
0    &  \gamma^{\nu}\gamma_5 X_{A[\alpha\beta]}^{[\gamma\delta]}\\  \end{array}\right)\cr
=&\Gamma_A  \left( \begin{array} {c c }
 \gamma^{\mu\nu\rho} \gamma_5   & 0    \\
0    &  \gamma^{\nu}\gamma_5 \\  \end{array}\right)= \left( \begin{array} {c c }
 \gamma^{\mu\nu\rho} \gamma_5   & 0    \\
0    &  \gamma^{\nu}\gamma_5 \\  \end{array}\right) \Gamma_A~~~.\cr
\end{align}

Rewriting Eq. \eqref{matrixM} to include the momentum argument $k$,
\begin{equation}\label{matrixM1}
{\cal M}\left(k\right)\left[\begin{array}{c}
 ~\gamma~~~\,\epsilon\kappa\\ \alpha\beta ~~\delta\\\end{array} \right] = \left( \begin{array} {c c }
-i  \gamma^{\mu\nu\rho} k_{\nu} \delta^\gamma_\delta  & -\gamma^{\mu} \overline{\phi}^{[\epsilon\kappa\gamma]}   \\
  \gamma^{\rho} \overline{\phi}^*_{[\alpha\beta\delta]}    &  -i\slashed{ k}\delta^{[\epsilon\kappa]}_{[\alpha\beta]} \\  \end{array}\right)~~~,
\end{equation}
we see that ${\cal M}(k+p) - {\cal M}(p)$ contains no off-diagonal matrix elements, and is given by
\begin{equation}\label{matrixM2}
{\cal M}\left(k+p\right)\left[\begin{array}{c}
 ~\gamma~~~\,\epsilon\kappa\\ \alpha\beta ~~\delta\\\end{array} \right]
-{\cal M}\left(p\right)\left[\begin{array}{c}
 ~\gamma~~~\,\epsilon\kappa\\ \alpha\beta ~~\delta\\\end{array} \right]
  = \left( \begin{array} {c c }
-i  \gamma^{\mu\nu\rho} k_{\nu} \delta^\gamma_\delta  & 0  \\
     0 &  -i\slashed{ k}\delta^{[\epsilon\kappa]}_{[\alpha\beta]} \\  \end{array}\right)~~~.
\end{equation}
Thus comparing with Eq. \eqref{vectorvertex}, we see that the vector vertex and the inverse propagator ${\cal M}={\cal N}^{-1}$ are related by the Ward identity
\begin{align}\label{ward1}
k_\nu {\cal V}^\nu_A=&i\Gamma_A  [{\cal M}(k+p)-{\cal M}(p)] = i[{\cal M}(k+p)-{\cal M}(p)]\Gamma_A\cr
=&i\Gamma_A  [{\cal N}^{-1}(k+p)-{\cal N}^{-1}(p)] = i[{\cal N}^{-1}(k+p)-{\cal N}^{-1}(p)]\Gamma_A~~~.\cr
\end{align}
The analogous Ward identity for the axial-vector vertex is obtained by right multiplication by $\gamma_5$.

\section{Anomaly calculation}

\subsection{Analysis for general symmetry-breaking $\overline{\phi}^{[\alpha\beta\gamma]}$ with $M$ invertible  }

We turn next to the anomaly calculation in the coupled model, closely following the methods of \cite{coupl}. Since the anomaly is a topological quantity, it
is independent of the magnitude of $\overline{\phi}$, so we can simplify the calculation by scaling $\overline{\phi}$ to infinity.  The propagator, vertex parts, and Ward identities   then become direct products of a Dirac matrix factor and an internal symmetry matrix factor, both of which have off-diagonal matrix elements equal to 0.  For the
upper diagonal matrix elements, the anomaly calculation then reduces to that of Sec. 13B of the abelianized case discussed in \cite{coupl}, times an internal symmetry factor.
For the lower diagonal matrix elements, the anomaly calculation reduces to that of Sec. 13A of \cite{coupl}, again times an internal symmetry factor.  The
combined result of the upper diagonal and lower diagonal contributions is
\begin{equation}\label{anomalyresult}
{\rm non-Abelian~coupled~model~anomaly}=(5D_{ABC} + {\cal D}_{ABC} ) \times {\rm standard~spin-\frac{1}{2}~anomaly}~~~,
\end{equation}
with upper diagonal contribution
\begin{equation}\label{DABCpart}
D_{ABC}= \frac{1}{2} {\rm Tr} t_A (t_B t_C+t_C t_B)
\end{equation}
the usual internal symmetry anomaly factor for a fundamental $8$ representation, and with lower diagonal contribution
\begin{equation}\label{calDABCpart}
{\cal D}_{ABC}=\frac{1}{2} {\rm Tr}PX_A(PX_B PX_C + PX_C  PX_B)
\end{equation}
an analogous internal symmetry factor for the $\overline{28}$ representation acted on by the projector $P$.
At first sight one worries that Eq. \eqref{calDABCpart} might give a non-integral anomaly, because of the
extra factors $P$ in the lower diagonal contribution, but since $P$ is a projector with diagonal
elements 0 or 1 this does not happen.  On the part of the $\overline{28}$ that couples
through $\overline{\phi}$ to the Rarita-Schwinger field and  which is already taken into account in the upper diagonal contribution
giving the $D_{ABC}$ term, $P$ acts as 0 and  there is no lower diagonal contribution.
On the orthogonal part of the $\overline{28}$ where $P$ acts as 1, which does not couple to the Rarita-Schwinger field, the lower diagonal
piece gives the usual group theoretic anomaly factor.

The factor of five   in the first term of Eq. \eqref{anomalyresult} corresponds precisely to the similar factor found in the
abelianized case.  A simple heuristic explanation for this factor is that, following   Alvarez-Gaum\'e and  Witten  \cite{lore},
the four-vector index $\mu$ of the Rarita-Schwinger field leads to an anomaly contribution that is a factor of 4 times the usual
spin-$\frac{1}{2}$ anomaly for the corresponding group representation.  For the uncoupled Rarita-Schwinger theory, there is
a Dirac first class constraint that requires gauge fixing and  ghosts, leading to an extra contribution of $-1$ times the
usual anomaly, giving the result of a factor of 3 times the usual anomaly cited in the papers of \cite{lore}.  However, in
the coupled model, the Rarita-Schwinger constraint is Dirac second class, which when exponentiated gives a non-propagating
ghost field with zero anomaly contribution.  Thus the total in this case is the naive 4 for the Rarita-Schwinger field, plus
an additional 1 for the spin-$\frac{1}{2}$ field to which it is coupled, giving the total of 5 found in \cite{coupl}, \cite{pais} for the
abelianized model and here, before inclusion of group theoretic factors,  for the non-Abelian case.  This heuristic argument also indicates that the way to restore
anomaly cancellation in the $SU(8)$ model of \cite{adler1} is to add an additional left chiral $\overline{8}$ spin-$\frac{1}{2}$ fermion field,
which makes an anomaly contribution of -1 times the usual spin-$\frac{1}{2}$ anomaly, compensating for the absence of a
similar contribution from ghost fields.

\subsection{Analysis for the specific symmetry breaking $\overline{\phi}$ produced by the Coleman-Weinberg mechanism}

In \cite{adler2} we analyzed the Coleman-Weinberg \cite{cole} symmetry breaking mechanism for the scalar field $\phi^{[\alpha\beta\gamma]}$ arising from its coupling to
$SU(8)$ gauge boson loops. We performed a numerical search of the parameter subspace that is compatible with the symmetry breaking chain $SU(8) \supset SU(3) \times G$, with $G$ to be
determined.  This gave a stable local minimum $\overline{\phi}^{[\alpha\beta\gamma]}$ with the only nonzero elements given by
\begin{align}\label{stablemin}
\overline{\phi}^{[123]}=&\overline{\phi}^{[178]}\equiv a=0.59762...~~~,\cr
\overline{\phi}^{[456]}\equiv & b=0.67199...~~~\cr
\end{align}
corresponding to the symmetry breaking chain $SU(8)\supset SU(3) \times Sp(4)$.  For this $\overline{\phi}$, the matrix $M$ defined in Eq. \eqref{Mdef} is
diagonal with nonzero diagonal elements
\begin{equation}\label{Mdiag}
M=2\,{\rm diag}(2|a|^2,|a|^2,|a|^2,|b|^2,|b|^2,|b|^2,|a|^2,|a|^2)~~~.
\end{equation}
Thus $M$ is invertible, and the analysis of  the previous section applies.  To see in detail how the anomaly is determined, we note that under  $SU(8)\supset SU(3) \times Sp(4)$, the $SU(8)$ fundamental $8$ representation decomposes as $8 =(1,1)+(1,4)+(3,1)$.   In terms of this branching decomposition,  the interaction Lagrangian density term
$\overline{\psi}_{\nu\gamma}\gamma^{\nu}\lambda_{[\alpha\beta]}\overline{\phi}^{[\alpha\beta\gamma]}$
in Eq. \eqref{covariant1} takes the form
\begin{equation}\label{interbreakdown}
\overline{\psi}_{\nu\gamma}\gamma^{\nu}\lambda_{[\alpha\beta]}\overline{\phi}^{[\alpha\beta\gamma]}=2(a \overline{\psi}_{\nu\, (1,1)}\gamma^\nu \lambda_{(1,1)}
+ a  \sum_{j=1}^4\overline{\psi}_{\nu\,(1,4)j}\gamma^\nu \lambda_{(1,4)j} + b \sum_{j=1}^3  \overline{\psi}_{\nu\,(3,1)j} \gamma^\nu \lambda_{(3,1)j})~~~,
\end{equation}
and similarly for its adjoint in Eq. \eqref{covariant1}. In Eq. \eqref{interbreakdown}, the first sum over $j$ goes over the four components of the 4 representation of
$Sp(4)$, and the second sum over $j$ goes over the three components of the 3 representation of $SU(3)$.   Under $SU(8)\supset SU(3) \times Sp(4)$, the $\overline{28}$ branches as
\begin{equation}\label{28bar}
\overline{28}= (\overline{3},4)+(1,5)+(\overline{3},1)+(1,1)+(1,4)+(3,1)~~~,
\end{equation}
so the pieces $(\overline{3},4)+(1,5)+(\overline{3},1)$ do not participate in the interaction term.  Therefore these pieces contribute standard spin-$\frac{1}{2}$ anomalies for each
representation,
whereas the pieces in Eq. \eqref{interbreakdown} together with their adjoints contribute a factor of 5 times the standard spin-$\frac{1}{2}$ anomalies for each representation.
To change the factor
of 5 to a factor of 4, which gives overall anomaly cancellation for the $SU(8)$ model, one adds a left chiral spin-$\frac{1}{2}$ representation $\overline{8}$ field, to give the
needed -1.
Since all representations of $Sp(4)$ are anomaly-free, the only anomaly contribution actually arising from  Eq. \eqref{interbreakdown} and its adjoint comes from the final term with coefficient $b$, which involves the 3 representation of $SU(3)$.  So anomaly cancellation consists of balancing the total number of $SU(3)$ triplet representations 3 against an equal number of $SU(3)$ anti-triplet representations $\overline{3}$.

\section{The extended $SU(8)$ model which guarantees anomaly cancellation and boson--fermion balance}

As we have seen in the preceding sections, to guarantee $SU(8)$ anomaly cancellation, the original model of \cite{adler1} must be extended to include an additional
representation $\overline{8}$ spin-$\frac{1}{2}$ fermion field.  This gives the left chiral fermion field content enumerated in Table I.

\begin{table} [ht]
\caption{Left chiral fermion field content of the extended $SU(8)$ model.  Square brackets indicate complete antisymmetrization of the enclosed indices.  The indices $\alpha,\beta,\gamma$ range from 1 to 8.}
\centering
\begin{tabular}{ c c c c c}
\hline\hline
field~~~ & spin~~~ & $SU(8)$ rep.~~~ & helicities~~~  \\
\hline
$\psi_{\mu}^{\alpha}$ & Weyl \,3/2 & 8 & 16  \\
$\lambda_{1[\alpha \beta]}$&Weyl\, 1/2&$\overline{28}$ & 56 \\
\hline
$\chi^{[\alpha\beta\gamma]}$&Weyl \, 1/2 & 56 & 112 \\
$\lambda_{2[\alpha \beta]}$&Weyl \,1/2&$\overline{28}$ & 56 \\
$\theta_\alpha$& Weyl \, 1/2& $\overline{8}$ & 16 \\
\hline\hline
\end{tabular}
\label{fieldcontentf}
\end{table}

\begin{table} [ht]
\caption{Boson field content of the extended $SU(8)$ model.  Square brackets indicate complete antisymmetrization of the enclosed indices.  The indices $\alpha,\beta,\gamma$ and $i$ range from 1 to 8, and the index $A$ runs from 1 to 63. }
\centering
\begin{tabular}{ c c c c c}
\hline\hline
field~~~ & spin~~~ & $SU(8)$ rep.~~~ & helicities~~~  \\
\hline
$h_{\mu \nu}$ & 2 & 1 & 2\\
$A_{\mu}^A$ & 1 & 63 & 126 \\
$\phi^{[\alpha\beta\gamma]}$& complex\, 0& 56 & 112 \\
$B_{\mu\,i}$&1 (each $i$)&1&16  \\
\hline\hline
\end{tabular}
\label{fieldcontentb}
\end{table}

We have denoted by $\lambda_{1[\alpha \beta]}$ the $\overline{28}$ spin-$\frac{1}{2}$ field that couples to the spin-$\frac{3}{2}$ field $\psi_{\mu}$ through an interaction
term analogous to that in Eq. \eqref{covariant1}, and have denoted by $\lambda_{2[\alpha \beta]}$ the orthogonal spin-$\frac{1}{2}$ field that does not couple
to the Rarita-Schwinger field.  The total helicity count for the fermion fields is 256.  The fields above the horizontal line, and the fields below the horizontal line,
separately cancel $SU(8)$ anomalies.  For the set of fields below the horizontal line, which corresponds to models previously studied by a number of authors \cite{su8models}, the $SU(8)$ anomaly count in units of the standard spin-$\frac{1}{2}$  anomaly   is $5$ for the $56$,  $-4$ for the $\overline{28}$, and $-1$ for the $\overline{8}$, totaling $0$.  For the fields above the horizontal line, the anomaly count is $4 \times 1=4$ for the coupled spin-$\frac{3}{2}$ field in representation 8,  which has a second class constraint, and $-4$ for the $\overline{28}$, again giving a total $SU(8)$ anomaly of 0.

In addition to anomaly cancellation, in \cite{adler1} where we introduced the $SU(8)$ model we required boson--fermion balance. That is, although full supersymmetry was not imposed, the numbers of fermion and
boson helicities were required to be equal.  Thus, to compensate for the added 16 helicities associated with the added $\overline{8}$ representation, it is necessary to add
16 helicities associated with bosonic degrees of freedom.  This can be done either by adding further gauge bosons, or by adding a complex scalar field in the $8$ or $\overline{8}$
representation.  For an added gauge boson to be compatible with the original $SU(8)$ gauging, its generators must commute with the $SU(8)$ generators, which by Schur's Lemma
requires them to be multiples of the unit $8 \times 8$ matrix.  This requires any added gauge boson to be a $U(1)$ gauging, and so to get the necessary 16 helicities the
added gauge group must be $U(1)^8$, with gauge fields $B_{\mu\,a},~a=1,...,8$. Since an analysis \cite{adler3} of paths to the Standard Model from the group $SU(3) \times Sp(4)$ shows that a missing $U(1)$ has
to be supplied, either dynamically or kinematically, we choose the option of adding a $U(1)^8$ gauge group over the alternative of adding a new scalar field.  This leads to extended boson sector of the
$SU(8)$ model given in Table II.

When there are $U(1)$ gauge fields,  conditions on their couplings are needed to guarantee anomaly cancellation.  Let $Y_i(r)$ denote the gauge coupling of the $i$th
$U(1)$ field to fermion representation $r$.  Cancellation of $SU(8)^2 U(1)$ triangle anomalies requires the condition
\begin{equation}\label{Ulinear}
\sum_r \ell(r)Y_i(r)=0
\end{equation}
for all $i$, with $l(r)$ the index \cite{slansky} of representation $r$.
Cancellation of the $U(1)^3$ triangle anomalies requires the condition
\begin{equation}\label{Ucubic}
\sum_r N(r) Y_i(r)Y_j(r)Y_k(r)=0
\end{equation}
for all $i,j,k$, with $N(r)$ the dimension of the representation $r$.  Since these conditions are homogeneous, we can
always choose the coupling to a given representation $r_1$ to be unity, that is $Y_i(r_1)=1$.  So  with 5 fermion representations
present in Table I, there are 4 relative couplings $Y_i(r)/Y_i(r_1)$ to be fixed by Eqs. \eqref{Ulinear} and \eqref{Ucubic} for each distinct $i$.

The simplest case is that in which all 8 $U(1)$ groups couple in the same way, so that no index $i$ is needed and we have
$Y_i(r)=Y(r)$.  There are then 2 conditions, and using the indices \cite{slansky}  $\ell(8)=1$, $\ell(28)=6$,  $\ell(56)=15$,  together
with $\ell(\overline{r})=\ell(r)$, and remembering the coupled model multiplier of 4 for the Rarita-Schwinger field, we have
\begin{align}\label{caseone}
0=&\sum_r \ell(r)Y(r)=4Y(8)+6Y(\overline{28}_1)+15Y(56)+6Y(\overline{28}_2)+Y(\overline{8})~~~,\cr
0=&\sum_r N(r) Y^3(r)=32Y^3(8)+ 28Y^3(\overline{28}_1)+56Y^3(56)+28Y^3(\overline{28}_2)+8Y^3(\overline{8})~~~.\cr
\end{align}
These conditions leave two of the four relative couplings as free parameters.  If we impose the additional constraint that $Y(8)=Y(\overline{28}_1)$, so
that the coupling term is a $U(1)$ invariant for fixed scalar field $\phi$, then only one relative coupling remains as a free parameter.

The next simplest case is that in which the 8 $U(1)$ groups form two sets $A$ and $B$, within which the $U(1)$ couplings are the same.
There are then 2 linear conditions (corresponding to $A$ and $B$ sums), and 4 cubic ones (corresponding to $A^3$, $B^3$, $A^2B$, $B^2A$ sums)
for a total of six in all.  Since there are 8 relative couplings, there are again 2 that are free parameters before imposing additional
restrictions.  When the 8 $U(1)$ groups form three or more sets,  within which the $U(1)$ couplings are the same,
there are in general more conditions than there are free relative couplings.\footnote{When there are $n$ such sets within which the $U(1)$ couplings are 
the same, there are $4n$ relative couplings to be fixed by the $2n+n(n-1)+n(n-1)(n-2)/6$ anomaly constraints.}

\section{Features of the revised model}

In this section we summarize some of the interesting features of the revised model, beyond the built-in anomaly cancellation and boson--fermion balance.

\begin{enumerate}
\item  Symmetry breaking through the rank three antisymmetric tensor $\phi^{[\alpha\beta\gamma]}$ gives a natural mechanism for getting an unbroken color $SU(3)$
subgroup.  When restricted to any subset of three values of the eight $SU(8)$  indices, the classical minimum $\bar{\phi}^{[\alpha\beta\gamma]}$ is proportional to the rank-three antisymmetric
tensor $\epsilon^{[\alpha\beta\gamma]}$ on those three index values, which is an $SU(3)$ invariant.  So no special tuning of the Higgs mechanism is needed to
get an unbroken color $SU(3)$ subgroup.
\item  The model is highly frustrated, in the sense that the chiral and $SU(8)$ symmetries prevent the appearance of bare mass terms in the Lagrangian for the
spin-$\frac{1}{2}$ fields.  Since all fields are left chiral, scalar bilinears of the form $\overline{\Psi}_{L1}\Psi_{L2}$ are automatically zero, so we need
only consider scalar bilinears of the form $\overline{\Psi}^c_{L1}\Psi_{L2}$, with the superscript $c$ denoting the charge conjugate field. Enumerating these, we have
(with $\lambda$ either $\lambda_1$ or $\lambda_2$),
\begin{align}\label{scalars}
&\bar{\theta}^c_{\alpha}\theta_{\beta}     \cr
&\bar{\theta}^c_{\alpha}\chi^{[\beta \gamma \delta]}    \cr
&\bar{\theta}^c_{\alpha} \lambda_{ [\beta \gamma]}    \cr
&\bar{\lambda}^c_{[\alpha \beta]} \lambda_{ [\gamma \delta]}    \cr
&\bar{\lambda}^c_{[\alpha \beta]}\chi^{[\gamma \delta \epsilon]}  \cr
&\bar{\chi}^{c\, [\alpha\beta\gamma]} \chi^{[\delta \epsilon \nu]}   \cr
\end{align}
In none of these can the indices be contracted to form an $SU(8)$ invariant, so no bare mass terms are allowed.  Only one of the above admits forming an $SU(8)$
invariant when contracted with the scalar field,
\begin{equation}\label{phiscalar}
\bar{\theta}^c_{\alpha} \lambda_{ [\beta \gamma]}  \phi^{[\alpha\beta\gamma]}~~~,
\end{equation}
but in the generic case, when $\theta$ and $\lambda$ do not have equal and opposite $U(1)$ charges,  this coupling will be forbidden by $U(1)$ gauge invariance.
So unless an additional restriction is imposed on the $U(1)$ charges,  we expect no spin-$\frac{1}{2}$ mass terms to appear through symmetry breaking by the scalar field $\phi$.  Thus fermion masses will have to be generated by dynamical symmetry breaking, above and beyond the symmetry breaking induced by the scalar field. This could
form the basis for giving a mass hierarchy, with calculable \cite{weinberg} spin-$\frac{1}{2}$  fermion masses.
\item Performing a similar enumeration with $\gamma^{\mu}\psi^{\alpha}_{\mu}$ replacing one of the spin-$\frac{1}{2}$ fields, we have for the
bilinears allowed by chiral symmetry,
\begin{align}\label{vectorscalar}
&\bar{\theta}^{\alpha}\gamma^{\mu}\psi^{\beta}_{\mu}   \cr
&\bar{\lambda}^{[\alpha \beta]}\gamma^{\mu}\psi^{\gamma}_{\mu} \cr
&\bar{\chi}_{[\alpha\beta\gamma]}\gamma^{\mu}\psi^{\delta}_{\mu}  \cr
&\bar{\psi}^{c \,\alpha}_{\nu}\gamma^{\nu} \gamma^{\mu}\psi^{\beta}_{\mu}   \cr
\end{align}
Again, in none of these can the indices be contracted to form an $SU(8)$ invariant, so no bare mass terms involving the spin-$\frac{3}{2}$ field are allowed.
Only one of the above admits forming an $SU(8)$
invariant when contracted with the scalar field,
\begin{equation}\label{phiscalar1}
\bar{\lambda}^{[\alpha \beta]}\gamma^{\mu}\psi^{\gamma}_{\mu} \phi^*_{[\alpha\beta\gamma]}~~~,
\end{equation}
which with its adjoint gives, after symmetry breaking, the coupling term introduced in Eq. \eqref{covariant1}.
\item  There is an interesting correspondence between the representation dimensions appearing in Tables I and II, and geometric features of the $E_8$ root lattice.
This may be pure coincidence, but could have a deep significance and indicate an origin of the $SU(8)$ model through a process on the
$E_8$ lattice.  The $E_8$ lattice is the set of points in $R^8$ with the properties \cite{wiki1}  that (i) all of the coordinates are integers or all of the
coordinates are half-integers, and (ii) the sum of the eight coordinates is an even integer.  It is generated by the 240  root vectors of $E_8$, which consist \cite{wiki2}
of the 112 roots with integer entries obtained from
\begin{equation}\label{integerroot}
(\pm 1,\pm 1,0,0,0,0,0,0)
\end{equation}
by taking an arbitrary combination of signs and an arbitrary permutation of
coordinates, and the 128 roots with half-integer entries obtained from
\begin{equation}
(\pm \frac{1}{2},\pm \frac{1}{2},\pm \frac{1}{2},\pm \frac{1}{2},\pm \frac{1}{2},\pm \frac{1}{2},\pm \frac{1}{2},\pm \frac{1}{2})
\end{equation}
by taking an even number of minus signs, so that the sum of all eight coordinates is even.  Representations $8$ or $\overline{8}$, with 16 helicities, have dimensions
corresponding to the numbers of positive and negative directions along the 8 $R^8$axes. The other dimensions appearing in Tables I and II relate to the geometry of the
$E_8$ lattice in more subtle ways.  The 2 helicities of the graviton and the 126 helicities of the $SU(8)$ gauge bosons involve numbers that connect to the geometry of the $E_8$ lattice
through the fact that any root and its negative, a set of 2, are orthogonal to exactly 126 other roots, irrespective of whether the initial 2
belong to the set of integer or half integer roots \cite{madore}.   The 112 real components of the representation 56 complex scalar $\phi$, and the 112 helicities of the representation 56 fermion $\chi$,  involve the same number as appears in the enumeration of the integer $E_8$ roots.  And the 56 helicities of the representation $\overline{28}$ fermions $\lambda_1$ or $\lambda_2$ involve numbers that connect to the geometry of the $E_8$ lattice through  the fact that any root makes an angle of $\pi/3$ with 56 others \cite{madore}. So all of the representation dimensions in the $SU(8)$ model connect with geometric properties of the $E_8$ roots and root lattice.  These intriguing relations warrant further investigation, to see if they can furnish an explanation for the mix of groups and representations assumed in the $SU(8)$ model.

\end{enumerate}

\end{document}